\documentclass{aa}
\usepackage[dvips]{graphicx}
\usepackage{psfig}
\newcommand{\ltsimeq}{\raisebox{-0.6ex}{$\,\stackrel{\raisebox{-.2ex}%
{$\textstyle<$}}{\sim}\,$}}

\begin{document}
 \title{New neighbours VI. Spectroscopy of DENIS nearby stars candidates}

   \author{F.~Crifo
       \inst{1}
       \and
       N.~Phan-Bao
       \inst{2,8}
       \and
       X.~Delfosse
       \inst{3}
       \and
       T.~Forveille
       \inst{4}
       \and
       J.~Guibert
       \inst{5}
       \and
       E.L. Mart\'{\i}n
       \inst{6,7}
       \and
	   C.~Reyl\'e
	   \inst{9}
       }

\offprints{F. Crifo, \email{francoise.crifo@obspm.fr}}

\institute{GEPI, Observatoire de Paris, 5 place J. Janssen, 92195 
           Meudon Cedex, France.
          \and
	  Institute of Astronomy and Astrophysics, Academia Sinica.
          P.O. Box 23-141, Taipei 106, Taiwan, R.O.C.
	  \and
	  Laboratoire d'Astrophysique de Grenoble, Universit\'e J. 
          Fourier, B.P. 53, F-38041 Grenoble, France.
	  \and
	  Canada-France-Hawaii Telescope Corporation, 65-1238 Mamalahoa 
           Highway, Kamuela, HI 96743 USA.
          \and
	  Centre d'Analyse des Images, GEPI, Observatoire de Paris,  
          61 avenue de l'Observatoire, 75014 Paris, France.
          \and
          Instituto de Astrof\'{\i}sica de Canarias, C/ V\'{\i}a L\'actea  
          s/n, E-38200 La Laguna (Tenerife), Spain. 
          \and
         University of Central Florida, Dept. of Physics, PO Box 162385, 
         Orlando, FL 32816-2385, USA.
          \and
         University of Hue, Dept. of Physics, 32 Le Loi, Hue, Vietnam.
          \and 	
         CNRS UMR 6091, Observatoire de Besan\c con, B.P. 15, 
           25010 Besan\c con Cedex, France.
		}

      \date{Received / Accepted}

\abstract{We present spectra of 36 nearby star candidates and 3 red giant
candidates, identified in the DENIS database by Phan-Bao et al. 
(\cite{phan-bao03}). 32 of the dwarf candidates are indeed nearby red 
dwarfs, with spectral types from M5.5 to M8.5. Out of 11 targets with 
low proper motion ($\mu < 0.1$~arc-sec.yr$^{\rm -1}$) but a 
Reduced Proper Motion above an inclusive threshold, 9 are red dwarfs.
The 4 contaminants are all reddened F--K main sequence stars, 
and could have been eliminated by checking for some well known high 
latitude molecular clouds. These stars might be of interest as probes of 
interstellar absorption.
For the red dwarfs we derive spectral types and spectroscopic distances, 
using a new calibration of the PC3 spectral index to absolute magnitudes 
in the $I, J, H$ and $K$ photometric bands.

We confirm 2 new members of the 12 pc volume (2 new M8.5), and one M7.5 NLTT 
object closer than 10pc;  and show that one quarter
of the stars with photometric distances under 30~pc have too small a proper 
motion for inclusion in the NLTT catalog.  
 }           
\titlerunning{New neighbours VI. Spectroscopic distances}
\authorrunning{F. Crifo et al.}
  \maketitle


\section{Introduction}

We present spectroscopic observations of candidate nearby red dwarfs
stars (d$<$30~pc; $2\leq I-J\leq3$)
from Phan-Bao et al. (\cite{phan-bao01} and 
\cite{phan-bao03}, hereafter Papers~I and II respectively). 
These late-M dwarf candidates were either found in the DENIS survey 
(Epchtein et al. \cite{epchtein97}) 
over 5700 square degrees, or cross-identified over a wider area between 
the DENIS database and the LHS or NLTT catalogs (Luyten \cite{luytena}, 
Luyten \cite{luytenb}). All were further selected by {"Maximum Reduced 
Proper Motion"} (hereafter MRPM, see Paper~II). That selection method 
retrieves solar neighbourhood red dwarfs down to very low proper motions, 
$\mu <$~0.1~arc-sec.yr$^{\rm -1}$, well below the 0.18~arc-sec.yr$^{\rm -1}$
threshold of the NLTT catalog (Luyten, 1979). 
Our current inventory of the immediate solar neighbourhood (particularly 
for faint members) leans heavily on the NLTT catalog and its 
predecessors, in spite of early efforts by Vyssotsky (1943; 1946; 1952) 
to correct this recognized kinematics bias through spectrophotometric 
selection. This makes low proper motion neighbours of particular interest,
and we felt that a spectroscopic validation of our selection criteria
was highly desirable.

Section 2 describes the sample and the spectroscopic observations. 
Section~3 presents the measurements of various spectroscopic indices
and derives an improved calibration of the PC3 spectral index 
(Mart\'{\i}n et al. \cite{martin99}) to absolute $I, J$, and $K$ magnitudes. 
It then
discusses the spectral types and distances of the confirmed late-M dwarfs. 
Section 4 discusses the contaminating giants and reddened earlier stars, 
examines the effectiveness of MRPM selection, and gives statistical elements
about the number of low-proper motion stars in the solar neighbourhood. 
The final section summarizes our results. 

\section{Data sample and spectroscopic observations}
\subsection{Data sample}
\label{Data_sample}

Papers~I and II examine a total of 132 DENIS sources with $2\leq I-J\leq3$,
and classify them into 80 probable dwarfs and 52 probable giants. The 52
giants and 62 of the dwarf candidates originate within 5700 square degrees,
where we examined all DENIS sources within well defined colour and
magnitude limits. They together constitute our main sample, which we use for
statistical discussions. 18 dwarf candidates were picked outside that area,
from a cross-identification of the DENIS database and the NLTT catalog. 
They constitute our extended sample, with a less direct statistical 
pedigree. The targets were divided into 3 priority groups, 
running from A (highest priority) to C (lowest):
\begin{itemize}
\item[\bf\Huge{.}] A: 11 probable red dwarfs with low proper motion (low-PM), 
  $\mu<$ 0.1 arc-sec.yr$^{\rm -1}$ (from Table~3 of Paper~II).
  Establishing their status and validating the selection method
  was our highest priority.

\item[\bf\Huge{.}] B and b: 69 very likely red dwarfs with high proper 
motion (high-PM), 
  $\mu >$ 0.1 arc-sec.yr$^{\rm -1}$ (from Table~5 of Paper~I and Tables 
  3a and 3b of Paper~II). 51 of them (noted B) belong to the main sample,
  and 18 (noted b) belong to the extended sample. 

\item[\bf\Huge{.}] C: 52 probable or previously known red giants (all 
  with low-PM, $\mu <$ 0.1 arc-sec.yr$^{\rm -1}$).
\end{itemize}

The A, B and C stars constitute the main sample, while the b targets
represent the extended sample. They were observed at ESO during two
observing runs, on the T152 and NTT telescopes.
\subsection{Spectroscopic observations and reductions}

\subsubsection{ T152 observations}
27 of the stars were observed with the Boller \& Chivens 
spectrograph mounted on the ESO 1.52m telescope at La Silla on August 
4$^{th}$, 5$^{th}$ and 6$^{th}$ 2002, shortly before that telescope was 
closed on October 1$^{st}$. The first night was cloudy and very windy, 
and we could only observe 2~targets, DENIS-P J0004$-$17 
and DENIS-P J0013$-$00, with mediocre seeing (1\arcsec.5). The two 
remaining nights were partly photometric and the seeing was around 
1\arcsec. The detector was La Silla CCD\#38 (2688$\times$512 pixels) and  
we used grating\#5, giving a dispersion of 0.13~nm~pixel$^{\rm -1}$ and
a wavelength coverage from 645~nm to 900~nm. We used the 2\arcsec slit, 
which corresponds to a spectral resolution of 0.32~nm~ for
the arc lines. The stellar images were narrower than the slit
width, and the stellar spectra therefore have somewhat higher spectral 
resolution. The small telescope size and a maximum exposure time of 55 minutes
prevented observation of the fainter targets. Low amplitude fringes are 
visible redwards of 760~nm~ in the raw spectra, but they flat-field 
out very well. We observed two spectrophotometric standards from
the ESO list (available at ftp.eso.org/pub/stecf/standards/ctiostan/),
EG~274 and LTT~7987.

We calibrated the spectra using standard IRAF packages: CCDPROC for 
bias-subtraction, flat-fielding and bad pixel correction, and APALL 
for wavelength and flux calibration, using He-Ar lamp spectra obtained 
at the beginning of each night and spectra of the spectrophotometric 
standards.

\subsubsection{NTT observations}
12 mostly fainter stars were observed at the NTT telescope in November 2003, 
using the EMMI intrument in its Red Imaging and Low-Dispersion mode (RILD).
In this mode the dispersion is 0.36 nm~pixel$^{\rm -1}$, and the effective 
wavelength range is 520 to 950 nm. The weather was photometric and the 
seeing  varied from 0.5 to 1.5 arcsec.
We selected the 1\arcsec~ slit, which corresponds to a spectral resolution 
of 1.04~nm~ for the arc lines.
Exposure times ranged from 30~s to 1200~s depending on the target's magnitude. 
The spectrophotometric standards, LTT~2415 and Feige~110 were chosen
from the ESO list.
All reduction was performed within MIDAS.

One more object was observed: DENIS P-J0410480-125142 $=$ LP 714-37. But as it 
was seen double with a large separation on the NTT pointing image, it is 
handled in a separate paper (Phan-Bao et al. 2005). No other visual binaries
were seen in the acquisition images.

\subsubsection{Analysis of the spectra}
All spectra were normalized to 1 over the 754-758~nm~ interval, 
the denominator of the PC3 index and a region with a good flat 
pseudo-continuum.

At the resolution of these spectra, the presence of the Na{\small I} 
and K{\small I}  doublets and the absence of the Ca{\small II} triplet 
immediately distinguish M dwarfs
from M giants, as does the general shape of the TiO bands. To our surprise,
a few of the targets also have much earlier spectra, reddened by 
intervening dust. One, J1625$-$24 (LP~862-26), had in fact been previously 
recognized as not an M dwarf (Reid \cite{reid03}, quoting a private 
communication by S.~Salim). Blue spectra would be needed for precise 
classification at these earlier spectral types, but the 4 stars are dwarfs 
of spectral type F to K.

The 39 observed stars divide into:
\begin{itemize}
\item[\bf\Huge{.}] 32 M dwarfs: 9 from group A, 18 from group B and 5 from
group b;
 
\item[\bf\Huge{.}] 3 giants, all from group C, as expected;

\item[\bf\Huge{.}] 4 highly reddened distant main-sequence stars, 2 from
group A and 1 each from groups B and b.
\end{itemize}
We could observe all group~A stars and therefore have a complete view of the 
low proper motion candidates.

\begin{figure}[h!]
\hspace{0.5cm}
\psfig{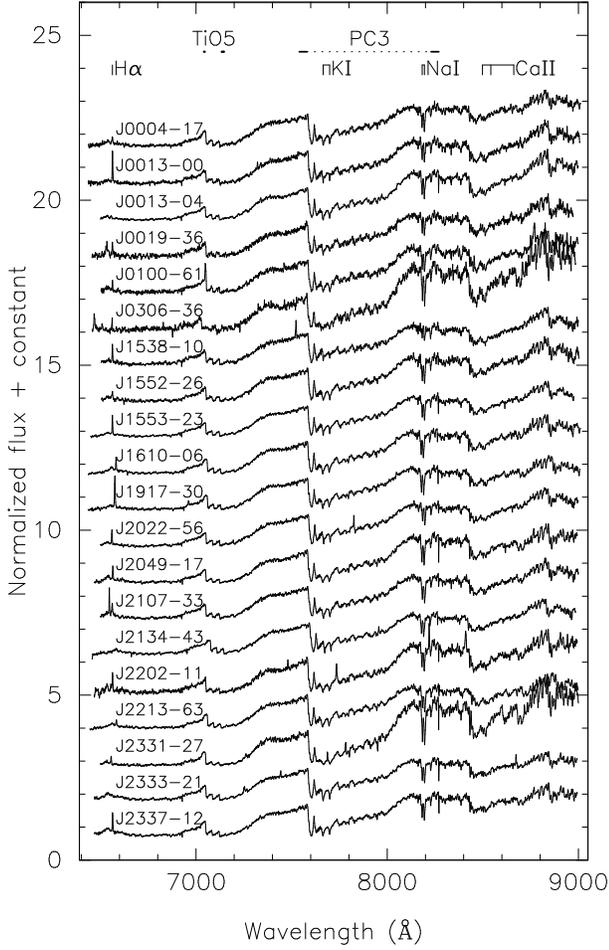}
\caption{Spectra of the 20 T152 late-M dwarfs. The positions of 
the H$_{\alpha}$,  Na{\small I}, K{\small I} and Ca{\small II} lines 
are indicated, as well as the spectral intervals used to compute the 
TiO5 and PC3 indices.}
\label{fig_spectre1}	
\end{figure}

\begin{figure}[h!]
\hspace{0.5cm}
\psfig{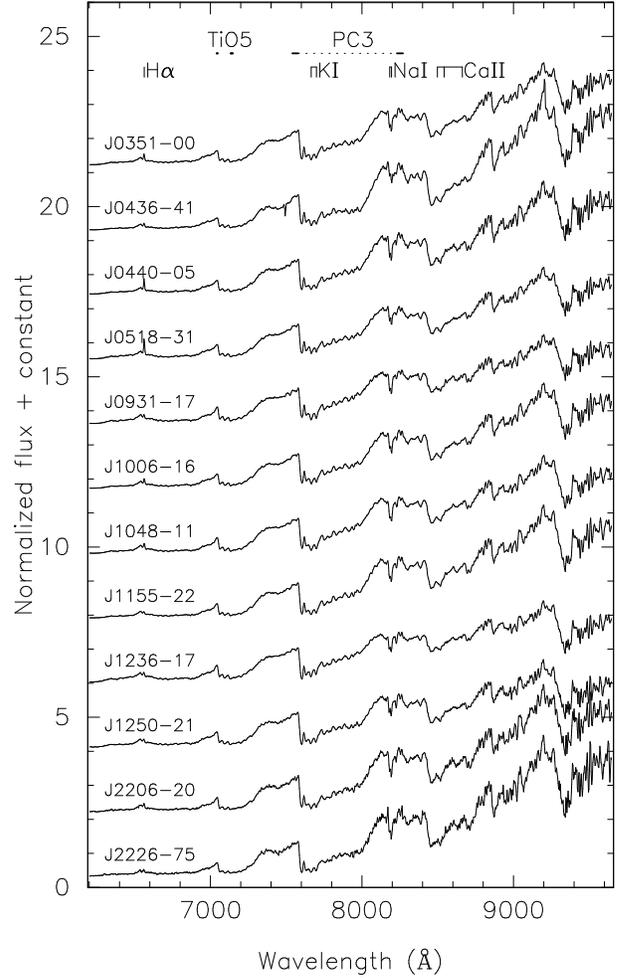}
\caption{Same as Figure 1, for the 12 NTT spectra.}
\label{fig_spectre2}	
\end{figure}

\begin{figure}[h!]
\hspace{0.5cm}
\psfig{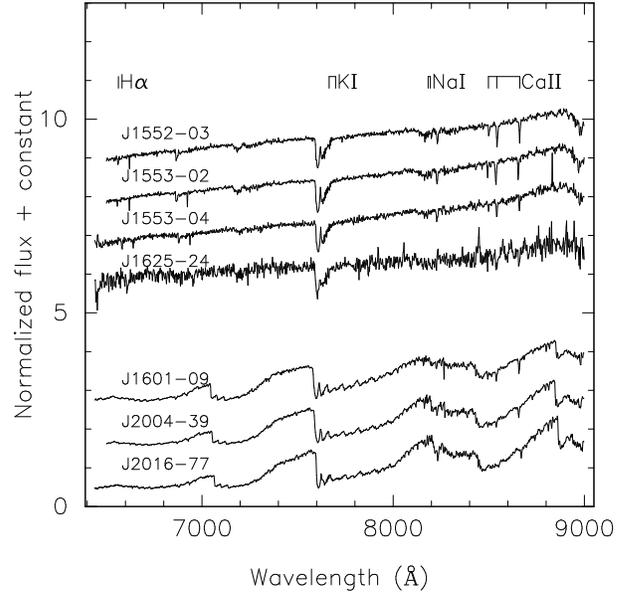}
\caption{Spectra of the three M giants rejected by the MRPM selection (bottom),
and the four reddened early-type main-sequence stars (top).}
\label{fig_spectre3}	
\end{figure}

Figures~\ref{fig_spectre1}, \ref{fig_spectre2} and \ref{fig_spectre3} 
respectively show the 20 T152 spectra and 12 NTT spectra of nearby 
late-M dwarfs and the 7 T152 spectra of distant stars. We defer a discussion 
of the later objects to Sec.~4.

%

%
\begin{figure}[h]
\psfig{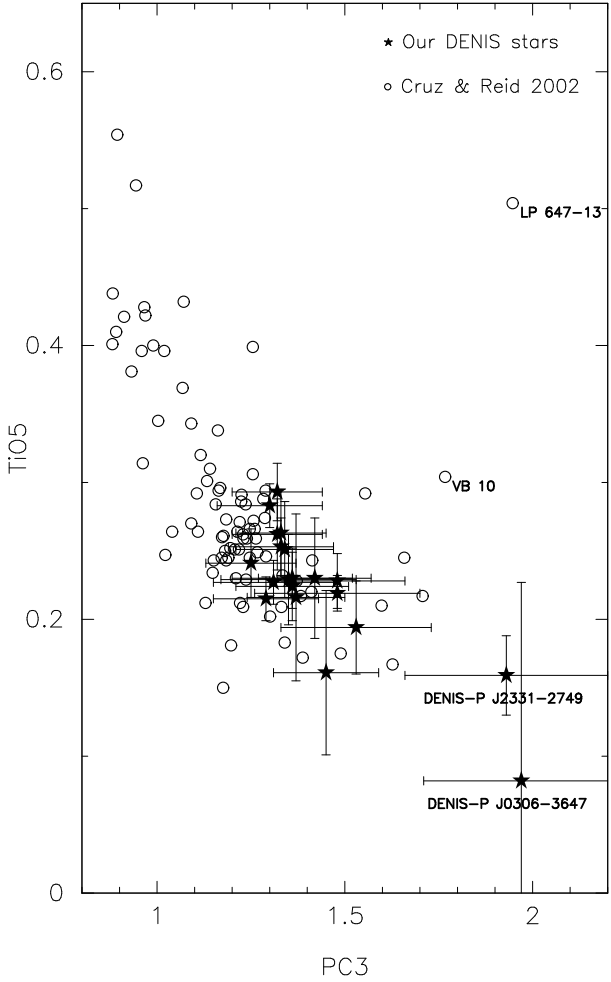}
\caption{TiO5 vs. PC3 diagram. The coolest stars are on the right side.
The filled stars represent the 20 T152
measurements from this paper, and the empty circles represent measurements
from Cruz \& Reid (\cite{cruz}). LP~647-13 (M9.0), VB~10 (flare star, 
M8.0), DENIS-P~J0306$-$3647 and DENIS-P~J2331$-$2749 (both M8.5 at 11~pc)
are individually identified. The error bars are determined from the 
flux fluctuations over the integration intervals, and therefore represent
lower limits to the true uncertainty. (The NTT stars are not plotted).
\label{fig_PC3_TiO5}
}
\end{figure}
%
%
\section{Spectroscopic indices and spectral type classification}
In the past few years, a number of interesting spectroscopic indices have 
been defined for M and L dwarfs (e.g. Reid et al. \cite{reid95}, 
Mart\'{\i}n et al. \cite{martin99}, Cruz \& Reid \cite{cruz}). L\'epine et al.
(\cite{lepine}) provides a good summary and selects the most effective
of those, which in combination allow to determine both spectral types
and qualitative metallicities. Table~\ref{table_indices} lists our 
measurements of the PC3, TiO5, CaH2, CaH3, VO1, VO2, TiO6 and TiO7 
spectroscopic indices for the 32 M dwarfs. PC3 measures the slope of 
a pseudo-continuum, and all other indices measure molecular features.
The low resolution of our NTT spectra biases their TiO5, CaH2 and CaH3 
indices in the direction of earlier spectral types, and those measurements 
should thus be interpreted with caution.

\begin{table*}
   \caption{Spectral indices for the 32  nearby red dwarfs observed at 
T152 and NTT}
 
    \label{table_indices}
  $$
   \begin{tabular}{lllllllllllll}
   \hline 
   \hline
   \noalign{\smallskip}
Stars              &  Other   & PC3   & TiO5     & CaH2  & CaH3   & VO1 & VO2 & TiO6   & TiO7  & SpT & SpT  & Pr\\
                   &   name      &                &       &        &     &     &       &     &       &  PC3   &  TiO5   &\\
  (1)&(2) &(3) &(4)  &(5) &(6) &(7) &(8)  &(9) &(10) & (11) & (12) & (13) \\
      \noalign{\smallskip}
\hline
T152 objects:      &              &        &         &         &         &         &         &         &         &        &       &   \\
J0004575$-$170937  &   ...        &  1.32  &  0.262  &  0.296  &  0.561  &  0.886  &  0.608  &  0.517  &  0.728  &  M5.5  &  M5.5 & B \\
J0013093$-$002551  &   ...        &  1.37  &  0.216  &  0.358  &  0.647  &  0.835  &  0.500  &  0.435  &  0.644  &  M5.5  &  M6.0 & A \\
J0013466$-$045736  &  LHS 1042    &  1.48  &  0.228  &  0.287  &  0.635  &  0.807  &  0.463  &  0.463  &  0.609  &  M6.0  &  M5.5 & B \\
J0019275$-$362015  &   ...        &  1.35  &  0.229  &  0.273  &  0.629  &  0.818  &  0.541  &  0.459  &  0.667  &  M5.5  &  M5.5 & B \\
J0100021$-$615627  &   ...        &  1.45  &  0.161  &  0.223  &  0.457  &  0.820  &  0.485  &  0.457  &  0.653  &  M6.0  &  M6.5 & A \\
J0306115$-$364753  &   ...        &  1.97  &  0.082  &  0.268  &  0.481  &  0.780  &  0.367  &  0.430  &  0.531  &  M8.5  &  M7.0:& B \\
J1538317$-$103850  &   ...        &  1.25  &  0.241  &  0.313  &  0.634  &  0.837  &  0.552  &  0.486  &  0.659  &  M5.0  &  M5.5 & A \\
J1552446$-$262313  &  LP 860-41   &  1.42  &  0.230  &  0.264  &  0.567  &  0.827  &  0.543  &  0.480  &  0.670  &  M6.0  &  M5.5 & b \\
J1553571$-$231152  &  LP 860-46   &  1.30  &  0.283  &  0.330  &  0.644  &  0.844  &  0.565  &  0.476  &  0.706  &  M5.0  &  M5.0 & b \\
J1610584$-$063132  &  LP 684-33   &  1.36  &  0.224  &  0.276  &  0.548  &  0.852  &  0.558  &  0.457  &  0.680  &  M5.5  &  M6.0 & B \\
J1917045$-$301920  &  LP 924-17   &  1.33  &  0.253  &  0.279  &  0.533  &  0.855  &  0.594  &  0.486  &  0.720  &  M5.5  &  M5.5 & b \\
J2022480$-$564556  &   ...        &  1.33  &  0.263  &  0.305  &  0.599  &  0.841  &  0.550  &  0.513  &  0.722  &  M5.5  &  M5.5 & A \\
J2049527$-$171608  &  LP 816-10   &  1.48  &  0.219  &  0.288  &  0.548  &  0.806  &  0.500  &  0.435  &  0.620  &  M6.0  &  M6.0 & B \\
J2107247$-$335733  &   ...        &  1.34  &  0.251  &  0.302  &  0.626  &  0.851  &  0.548  &  0.495  &  0.692  &  M5.5  &  M5.5 & B \\
J2134222$-$431610  &  WT 792      &  1.32  &  0.293  &  0.292  &  0.592  &  0.833  &  0.574  &  0.531  &  0.712  &  M5.5  &  M5.0 & B \\
J2202112$-$110945  &  LP 759-17   &  1.53  &  0.194  &  0.302  &  0.596  &  0.769  &  0.453  &  0.397  &  0.520  &  M6.5  &  M6.0 & B \\
J2213504$-$634210  &  WT 887      &  1.29  &  0.215  &  0.237  &  0.488  &  0.856  &  0.576  &  0.464  &  0.733  &  M5.0  &  M6.0 & B \\
J2331217$-$274949  &   ...        &  1.93  &  0.159  &  0.207  &  0.453  &  0.848  &  0.404  &  0.376  &  0.560  &  M8.5  &  M7.0:& B \\
J2333405$-$213353  &  LHS 3970    &  1.31  &  0.227  &  0.306  &  0.620  &  0.860  &  0.577  &  0.506  &  0.699  &  M5.0  &  M6.0 & B \\
J2337383$-$125027  &  LP 763-3    &  1.36  &  0.230  &  0.307  &  0.584  &  0.847  &  0.562  &  0.489  &  0.686  &  M5.5  &  M5.5 & B \\
                   &              &        &         &         &         &         &         &         &         &        &       &   \\
\hline
NTT objects:       &              &        &         &         &         &         &         &         &         &        &       &   \\
J0351000$-$005244  &  GJ 3252     &  1.62  &  0.251: &  0.307: &  0.654: &  0.745  &  0.437  &  0.451  &  0.592  &  M7.0  &  M5.5:& B \\
J0436278$-$411446  &   ...        &  1.88  &  0.226: &  0.282: &  0.683: &  0.648  &  0.308  &  0.402  &  0.469  &  M8.0  &  M7.5:& A \\
J0440231$-$053009  &  LP 655-48   &  1.75  &  0.247: &  0.300: &  0.603: &  0.750  &  0.422  &  0.430  &  0.569  &  M7.5  &  M7.5:& b \\
J0518113$-$310153  &   ...        &  1.49  &  0.216: &  0.273: &  0.602: &  0.755  &  0.448  &  0.413  &  0.597  &  M6.5  &  M6.0:& A \\
J0931223$-$171742  &  LP 788-1    &  1.53  &  0.224: &  0.264: &  0.534: &  0.808  &  0.531  &  0.472  &  0.671  &  M6.5  &  M6.0:& b \\
J1006319$-$165326  &  LP 789-23   &  1.66  &  0.248: &  0.301: &  0.626: &  0.743  &  0.429  &  0.448  &  0.585  &  M7.0  &  M5.5:& B \\
J1048126$-$112009  &  GJ 3622     &  1.60  &  0.183: &  0.227: &  0.521: &  0.787  &  0.475  &  0.437  &  0.617  &  M7.0  &  M6.0:& B \\
J1155429$-$222458  &  LP 851-346  &  1.75  &  0.199: &  0.248: &  0.546: &  0.753  &  0.423  &  0.421  &  0.557  &  M7.5  &  M7.5:& B \\
J1236396$-$172216  &   ...        &  1.43  &  0.251: &  0.291: &  0.568: &  0.822  &  0.591  &  0.524  &  0.693  &  M6.0  &  M5.5:& A \\
J1250526$-$212113  &   ...        &  1.50  &  0.214: &  0.291: &  0.640: &  0.748  &  0.435  &  0.434  &  0.547  &  M6.5  &  M6.0:& B \\
J2206227$-$204706  &   ...        &  1.81  &  0.209: &  0.265: &  0.511: &  0.745  &  0.398  &  0.411  &  0.545  &  M8.0  &  M7.5:& A \\
J2226443$-$750342  &   ...        &  1.91  &  0.290: &  0.329: &  0.600: &  0.733  &  0.379  &  0.452  &  0.524  &  M8.5  &  M8.0:& A \\

    \noalign{\smallskip}
    \hline 
   \end{tabular}
  $$

  \begin{trivlist}
  \item[] 
Columns 1 \& 2 : Full DENIS name and other name.

 \item[] 
Columns 3-10: Spectroscopic indices. PC3 is defined in Mart\'{\i}n et al. 
(\cite{martin99}); TiO5, CaH2 and CaH3 in Reid et al. (\cite{reid95});
VO1 in Hawley et al. (\cite{hawley02}); VO2, TiO6 and TiO7 in L\'epine 
et al. (\cite{lepine}).

 \item[] 
Columns 11 \& 12: Spectral types derived from the (PC3, spectral type) 
relation of Mart\'{\i}n et al. (\cite{martin99}) and from the 
(TiO5, spectral type) relation of Cruz \& Reid (\cite{cruz}), and rounded
to the nearest half type. The types followed by ":" are 
unreliable, as discussed in the text, Sec. 2.1, as well as the NTT indices
TiO5, CaH2 and CaH3 .

 \item[] 
Column 13: Priority group (A, B or b), as defined in Sec. 2.1.

  \end{trivlist}

\end{table*}

Two of the indices, PC3 and TiO5, are of particular value for 
spectral type determination, while the others mostly provide
information on the metallicity. PC3 and TiO5 have complementary
strengths and weaknesses:
\begin{itemize}
\item[\bf\Huge{.}] The defining bands of PC3 (numerator: 823--827~nm, 
denominator: 754--758~nm, Mart\'{\i}n et al. \cite{martin99}) are wide, 
but separated by almost 80~nm. The index is therefore easily measured 
with good signal to noise ratio, immune to moderate changes of the spectral 
resolution, and insensitive to small errors in the wavelength calibration. It 
is on the other hand rather sensitive to an imperfect relative flux 
calibration.
On the astrophysical side, PC3 is an excellent spectral type indicator
over the [M2.5, L1] range, and it increases monotonically over the full
M class. 
\item[\bf\Huge{.}] The two defining bands of TiO5 (numerator: 
712.6--713.5~nm, 
denominator: 704.2--704.6~nm, Reid et al. \cite{reid95}, Cruz \& Reid 
\cite{cruz}) are within less than 10~nm, and the index is therefore
immune to flux calibration errors. On the other hand, the interval for
the denominator is narrow and positioned at the very head of a molecular 
band. The index is therefore sensitive to both spectral resolution 
changes and small errors in the wavelength scale. As the denominator 
contains few data points, a good measurement of TiO5 also requires a 
spectrum with significantly higher signal to noise ratio than PC3. On 
the astrophysical side, TiO5 is an excellent spectral type diagnostic 
for early M dwarfs, but it saturates and wraps around at spectral type 
M7. For late subtypes (e.g. J0306115$-$364753) the flux in the TiO5 
numerator is also very low and the index therefore has large error bars
except for very high S/N spectra. It is therefore of limited value beyond
$\sim$M6.
 
In our case, the resolution of NTT spectra is too low ($\sim$10\AA); the
TiO5 index is therefore not taken into account for them.
\end{itemize} 

Figure~\ref{fig_PC3_TiO5} compares the PC3 and TiO5 indices for our T152
spectra, and demonstrates the saturation of TiO5 at late spectral types. 
Since our spectra have moderate signal to noise ratios but good relative flux 
calibration, we use the PC3 index as our main classification tool. We adopt 
the (Mart\'{\i}n et al. \cite{martin99}) calibration of the PC3 index to 
spectral type, to determine the spectral class with an uncertainty 
of $\pm$0.5 subclass.

For three new objects, DENIS-P J0306115$-$364753, 
DENIS-P J2226443$-$750342 and DENIS-P J2331217$-$274949 
(hereafter DENIS-P J0306$-$3647, DENIS-P J2226$-$7503 and 
DENIS-P J2331$-$2749), the PC3 index indicates an M8.5$\pm$0.5 
spectral type. Their spectral types from TiO5 would be about M7, but 
the 3 spectra are evidently later than M7. The stars are therefore 
simply beyond the validity range of the TiO5 calibration. Two of 
them are close neighbours of the Sun, both at 11$\pm$1.2~pc 
(Section~4) and clearly of interest for parallax programs. 

The CaH and VO indices (not plotted) do not reveal any subdwarfs. This is
expected since our sample is mostly volume-limited, and too small to have
much probability of including members of the stellar halo.

\subsection{PC3 index-magnitude relation}
%
%
\begin{figure}
\psfig{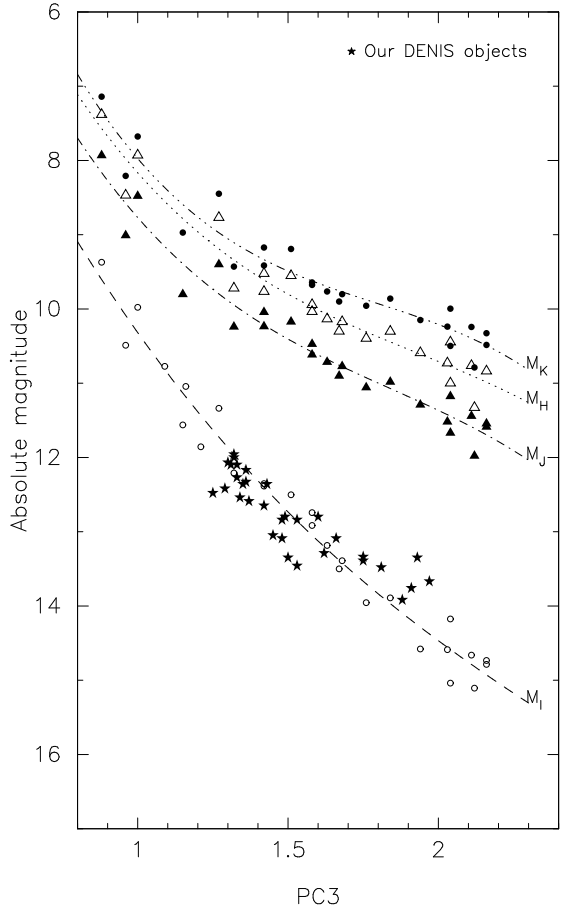}
\caption{Absolute magnitudes in the Cousins $I$ and CIT $JHK$ passbands 
as a function
of PC3 index (data in Table~\ref{table_cali}).The $M_{\rm I}$ values used for 
our objects are the ones derived from the DENIS ($I-J$) photometry in Papers
I and II and given again in Table 3, col.3. 
}
\label{fig_index_mag}
\end{figure}

We identified in the literature 27 stars with good trigonometric 
parallaxes and photometry (Table~\ref{table_cali})
and with PC3 indices from Mart\'{\i}n et al. (\cite{martin99}), Geballe 
et al. (\cite{geballe}) and Cruz \& Reid (\cite{cruz}), allowing to derive a
calibration of absolute magnitudes in $I$, $J$, $H$, $K$ versus PC3 index 
(Phan-Bao \cite{phan-bao02} contains an earlier version).

We transformed photometry from Leggett et al. (\cite{leggett98}) and Weis 
(\cite{weis}) to the Cousins-CIT systems, using the colour transformations 
published in the two papers. For our present purpose the DENIS system
is close enough to Cousins-CTIO ($\ltsimeq$0.05~mag, Delfosse 
\cite{delfosseb}), and we therefore use the DENIS photometry without 
any corrections.

\begin{table*}
   \caption{Red dwarfs with PC3 index, Cousins-CIT photometry and 
trigonometric parallaxes, used for the absolute magnitude calibration}
    \label{table_cali}
  $$
   \begin{tabular}{llllllllll}
   \hline 
   \hline
   \noalign{\smallskip}
Stars &$\alpha_{\rm 2000}$, $\delta_{\rm 2000}$&    Sp.T.  & PC3 & $I$  &$I-J$ & $H$  & $K$  & $\pi$ & References \\ 
  (1)&(2) &(3) &(4)  &(5) &(6) &(7) &(8)  &(9) &(10) \\
   \hline
   \noalign{\smallskip}  
LHS 4           &  00 18 25.79  $+$44 01 38.2  &  M3.5 &  0.96  &~\,8.25 &  1.48  &~\,6.23 &~\,5.97 &  280.3$\pm$1.1 &  2, 6, 6, 9, 9, 9, 9, 16      \\                           
LP 292-67       &  00 20 29.50  $+$33 05 05.5  &  M5.5 &  1.21  & 12.36  &~\,...  &~\,...  &~\,...  &~\,79.3$\pm$3.7 &  2, 8, 7, 15,...,...,..., 17  \\
RG 0050-2722    &  00 52 54.67  $-$27 05 59.5  &  M9   &  2.16  & 16.67  &  3.19  &~\,...  & 12.42  &~\,41.0$\pm$4.0 &  1, 5, 5, 1, 1,..., 1, 17     \\    
CTI 0126+28     &  01 27 39.13  $+$28 05 54.4  &  M9   &  2.11  & 17.24  &  3.22  & 13.34  & 12.82  &~\,30.5$\pm$0.5 &  4, 5, 5, 11, 11, 11, 11, 11   \\ 
LHS 17          &  02 46 14.97  $-$04 59 21.5  &  M6   &  1.15  & 12.66  &  1.76  &~\,...  & 10.07  &~\,60.3$\pm$8.2 &  1, 8, 7, 1, 1,..., 1, 17     \\
LP 412-31       &  03 20 59.70  $+$18 54 22.7  &  M8   &  1.84  & 14.70  &  2.91  & 11.11  & 10.67  &~\,68.9$\pm$0.6 &  3, 5, 5, 11, 11, 11, 11, 11  \\                           
LHS 2065        &  08 53 36.11  $-$03 29 32.4  &  M9   &  2.16  & 14.44  &  3.20  & 10.49  &~\,9.98 &  117.3$\pm$1.5 &  2, 5, 5, 9, 9, 9, 9, 17      \\                           
LHS 292         &  10 48 12.64  $-$11 20 09.8  &  M6.5 &  1.58  & 11.20  &  2.30  &~\,8.32 &~\,7.96 &  220.3$\pm$3.6 &  1, 6, 6, 9, 9, 9, 9, 17       \\                           
GJ 406          &  10 56 28.99  $+$07 00 52.0  &  M6   &  1.51  &~\,9.39 &  2.33  &~\,6.44 &~\,6.08 &  419.1$\pm$2.1 &  2, 5, 5, 9, 9, 9, 9, 17       \\   
LHS 39          &  11 05 30.31  $+$43 31 16.6  &  M5.5 &  1.32  & 10.63  &  1.97  &~\,8.14 &~\,7.85 &  206.9$\pm$1.2 &  2, 6, 6, 9, 9, 9, 9, 16       \\                          
LHS 2397a       &  11 21 49.21  $-$13 13 08.3  &  M8.5 &  2.04  & 14.95  &  3.00  & 11.22  & 10.77  &~\,70.0$\pm$2.1 &  1, 5, 5, 9, 9, 9, 9, 17        \\                         
BRI 1222-1221   &  12 24 52.28  $-$12 38 35.2  &  M8.5 &  1.94  & 15.74  &  3.29  & 11.75  & 11.31  &~\,58.6$\pm$3.8 &  1, 5, 5, 10, 10, 10, 10, 13   \\
LHS 330         &  12 29 14.28  $+$53 33 05.0  &  M6   &  1.42  & 14.38  &  2.15  & 11.76  & 11.41  &~\,39.9$\pm$1.0 &  2, 6, 6, 9, 9, 9, 9, 17       \\                          
LHS 2924        &  14 28 43.33  $+$33 10 37.9  &  M9   &  2.04  & 15.21  &  3.37  & 11.17  & 10.67  &~\,92.4$\pm$1.3 &  2, 5, 5, 9, 9, 9, 9, 17       \\                           
LHS 2930        &  14 30 37.95  $+$59 43 24.1  &  M6.5 &  1.68  & 13.31  &  2.62  & 10.09  &~\,9.72 &  103.8$\pm$1.3 &  2, 6, 6, 9, 9, 9, 9, 17       \\           
LHS 3003        &  14 56 38.38  $-$28 09 49.7  &  M7   &  1.67  & 12.53  &  2.60  &~\,9.33 &~\,8.93 &  156.3$\pm$0.3 &  2, 6, 6, 9, 9, 9, 9, 17       \\
T* 513-46546    &  15 01 08.19  $+$22 50 02.5  &  M9   &  2.12  & 15.09  &  3.13  & 11.31  & 10.77  &  100.8$\pm$2.3 &  4, 5, 5, 12, 12, 12, 12, 17   \\                           
T* 868-110639   &  15 10 16.86  $-$02 41 07.4  &  M9   &  2.03  & 15.79  &  3.07  & 11.93  & 11.44  &~\,57.5$\pm$1.9 &  1, 5, 5, 12, 12, 12, 12, 17   \\                            
LHS 427         &  16 55 25.32  $-$08 19 21.9  &  M3.5 &  1.00  &~\,9.04 &  1.50  &~\,6.99 &~\,6.74 &  154.0$\pm$4.0 &  2, 6, 6, 9, 9, 9, 9, 16      \\                           
VB 8            &  16 55 35.35  $-$08 23 42.3  &  M7   &  1.63  & 12.24  &  2.47  &~\,9.19 &~\,8.82 &  154.5$\pm$0.7 &  2, 8, 7, 9, 9, 9, 9, 18       \\                   
LHS 3339        &  17 55 33.50  $+$58 24 26.5  &  M6   &  1.42  & 14.02  &  2.31  & 11.19  & 10.84  &~\,46.4$\pm$1.0 &  2, 6, 6, 10, 10, 10, 10, 17  \\
GJ 1224         &  18 07 32.88  $-$15 57 46.8  &  M4.5 &  1.16  & 10.43  &~\,...  &~\,...  &~\,...  &  132.6$\pm$3.7 &  2, 8, 7, 15,...,...,..., 17   \\          
GJ 1227         &  18 22 27.28  $+$62 03 00.8  &  M4.5 &  1.09  & 10.35  &~\,...  &~\,...  &~\,...  &  121.5$\pm$2.2 &  2, 8, 7, 14,...,...,..., 17   \\       
Gl 720B         &  18 35 27.39  $+$45 45 39.6  &  M3.5 &  0.88  & 10.32  &  1.44  &~\,8.33 &~\,8.09 &~\,64.6$\pm$1.0 &  2, 8, 7, 9, 9, 9, 9, 16       \\                            
VB 10           &  19 16 57.66  $+$05 09 00.4  &  M8   &  1.76  & 12.80  &  2.90  &~\,9.24 &~\,8.80 &  170.3$\pm$1.4 &  2, 5, 5, 9, 9, 9, 9, 16       \\                           
LHS 523         &  22 28 54.38  $-$13 25 17.8  &  M6.5 &  1.58  & 13.00  &  2.27  & 10.20  &~\,9.90 &~\,88.8$\pm$4.9 &  1, 6, 6, 9, 9, 9, 9, 17       \\                           
GJ 905          &  23 41 55.17  $+$44 10 38.0  &  M5   &  1.27  &~\,8.84 &  1.94  &~\,6.27 &~\,5.95 &  316.0$\pm$1.1 &  2, 5, 5, 9, 9, 9, 9, 17       \\                           
    \noalign{\smallskip}
    \hline 
   \end{tabular}
  $$
  \begin{list}{}{}
  \item[T*] : TVLM 

Columns 1 \& 2: Object name and coordinates for equinox and epoch J2000.

Columns 3 \& 4: Spectral type and PC3 index.

Columns 5, 6, 7 \& 8: Optical and infrared photometry in the 
Cousins-CIT systems.

Column 9:  Trigonometric parallax and its standard error, in mas. 

Column 10: References for all parameters listed in the columns 2-9, column by
column:
1) DENIS; 2) Bakos et al. \cite{bakos}; 3) Salim \& Gould \cite{salim}; 
4) Monet et al. \cite{monet03}; 5) Mart\'{\i}n et al. \cite{martin99}; 
6) Geballe et al. \cite{geballe}; 7) Cruz \& Reid \cite{cruz}; 8) Hawley 
et al. \cite{hawley96}; 9) Leggett \cite{leggett}; 
10) Leggett et al. \cite{leggett98}; 11) Dahn et al. \cite{dahn02}; 12) 
Tinney et al. \cite{tinneya}; 
13) Tinney \cite{tinney96}; 14) Bessell \cite{bessell}; 
15) Weis \cite{weis}; 16) HIP (ESA \cite{esa}); 
17) GCTP (van Altena et al. \cite{vanaltena}); 18) Monet et al. 
\cite{monet92}.
  \end{list}
\end{table*}
We excluded known doubles, unless the separation of the two stars is sufficient
for individual photometric observations (e.g. Gl~720B, vB8, vB10), and all
large amplitude variables. We did however have to accept a number of 
low amplitude flare stars, with peak visible light amplitude of 0.1 to 
0.3~mag, since many very late M dwarfs have some photometric variability.

Table~\ref{table_cali} summarizes the information, and 
Fig.~\ref{fig_index_mag} shows that the PC3 index tightly maps
to absolute magnitudes in either of the $I$, $J$, $H$, and $K$ bands.
The following cubic least-square fits to those data:
\begin{eqnarray}
M_{\rm I} & = & 2.227+10.871(PC3)-3.185(PC3)^{2} \nonumber \\
           &   & +0.4048(PC3)^{3} \label{eq1} \\
M_{\rm J} & = & -0.5630+15.384(PC3)-7.396(PC3)^{2} \nonumber \\
           &   & +1.343(PC3)^{3} \label{eq2} \\ 
M_{\rm H} & = & -0.7493+14.408(PC3)-6.647(PC3)^{2} \nonumber \\
           &   & +1.154(PC3)^{3} \label{eq3} \\
M_{\rm K} & = & -2.877+18.779(PC3)-9.738(PC3)^{2} \nonumber \\
           &   & +1.810(PC3)^{3} \label{eq4} 
\end{eqnarray}

are valid for $0.9 \leq PC3 \leq 2.2$, or spectral types between M3.5
and M9. Over this range the rms dispersion of the data around
the fits is approximately 0.25~magnitude, in any of the 4 photometric 
bands. This corresponds to a $\sim$12\% standard error on distances
to single stars.

\subsection{Spectral types and Distances}
\begin{table*}
   \caption{Spectrophotometric distance and spectral type for 
    the 32 nearby M dwarfs}
    \label{table_resultat}
  $$
   \begin{tabular}{lllllllllllllll}
   \hline 
   \hline
   \noalign{\smallskip}
Stars     & $I$     & $I-J$ & $J-K$   &PC3  & $M_{I}$   & $M_{I}$ & $M_{J}$ & $M_{K}$ & $d_{I}$  & $d_{J}$ & $d_{K}$  & $d_{sp}$ & $d$   & Sp. \\
          &         &       &         &     &   ($IJ$)   &  (PC3)  & (PC3)   &   (PC3) &  (PC3) &  (PC3)  &   (PC3)  &  (PC3)   &  ($IJ$)     &type  \\
  (1)&(2) &(3) &(4)  &(5) &(6) &(7) &(8)  &(9) &(10) & (11) & (12) & (13) & (14) & (15) \\
   \hline
   \noalign{\smallskip}  
J0004-17  &  13.00  &  2.03  &  0.93  &  1.32  &  12.00  &  11.96  &~\,9.95  &~\,9.11  &  16.2  &  16.0  &  15.4  &  15.9  &  15.9  &  M5.5   \\
J0013-00  &  14.37  &  2.22  &  0.88  &  1.37  &  12.59  &  12.18  &  10.08  &~\,9.23  &  27.4  &  25.9  &  25.6  &  26.3  &  22.7  &  M5.5   \\
J0013-04  &  13.88  &  2.44  &  0.99  &  1.48  &  13.09  &  12.65  &  10.36  &~\,9.45  &  17.6  &  16.5  &  15.8  &  16.6  &  14.4  &  M6.0   \\
J0019-36  &  14.30  &  2.14  &  0.89  &  1.35  &  12.36  &  12.09  &  10.03  &~\,9.18  &  27.6  &  26.7  &  26.2  &  26.8  &  24.4  &  M5.5   \\
J0100-61  &  15.01  &  2.42  &  0.94  &  1.45  &  13.05  &  12.53  &  10.29  &~\,9.40  &  31.4  &  28.9  &  28.2  &  29.5  &  24.6  &  M6.0   \\
J0306-36  &  14.41  &  2.79  &  1.03  &  1.97  &  13.67  &  14.38  &  11.31  &  10.16  &  10.2  &  11.5  &  12.2  &  11.3  &  14.0  &  M8.5   \\
J0351-00  &  13.75  &  2.55  &  0.99  &  1.62  &  13.29  &  13.20  &  10.66  &~\,9.68  &  12.9  &  12.8  &  12.7  &  12.8  &  12.4  &  M7.0   \\
J0436-41  &  16.04  &  2.92  &  1.12  &  1.88  &  13.92  &  14.10  &  11.14  &  10.04  &  24.5  &  24.9  &  24.7  &  24.7  &  26.6  &  M8.0   \\
J0440-05  &  13.35  &  2.61  &  1.19  &  1.75  &  13.39  &  13.67  &  10.91  &~\,9.86  &~\,8.6  &~\,9.3  &~\,8.7  &~\,8.9  &~\,9.8  &  M7.5   \\
J0518-31  &  14.17  &  2.30  &  1.00  &  1.49  &  12.80  &  12.69  &  10.38  &~\,9.47  &  19.7  &  19.8  &  19.0  &  19.5  &  18.8  &  M6.5   \\
J0931-17  &  13.36  &  2.32  &  1.01  &  1.53  &  12.84  &  12.85  &  10.47  &~\,9.54  &  12.6  &  13.0  &  12.5  &  12.7  &  12.7  &  M6.5   \\
J1006-16  &  14.55  &  2.44  &  1.14  &  1.66  &  13.09  &  13.35  &  10.74  &~\,9.74  &  17.4  &  18.8  &  17.6  &  17.9  &  19.6  &  M7.0   \\
J1048-11  &  11.25  &  2.30  &  0.98  &  1.60  &  12.80  &  13.13  &  10.62  &~\,9.65  &~\,4.2  &~\,4.6  &~\,4.6  &~\,4.5  &~\,4.9  &  M7.0   \\
J1155-22  &  13.48  &  2.58  &  1.05  &  1.75  &  13.34  &  13.67  &  10.91  &~\,9.86  &~\,9.2  &  10.0  &~\,9.9  &~\,9.7  &  10.7  &  M7.5   \\
J1236-17  &  13.91  &  2.14  &  1.14  &  1.43  &  12.36  &  12.44  &  10.24  &~\,9.36  &  19.6  &  20.2  &  18.0  &  19.3  &  20.4  &  M6.0   \\
J1250-21  &  13.78  &  2.59  &  1.11  &  1.50  &  13.35  &  12.73  &  10.40  &~\,9.49  &  16.2  &  14.4  &  13.1  &  14.6  &  12.2  &  M6.5   \\
J1538-10  &  14.36  &  2.18  &  0.95  &  1.25  &  12.48  &  11.63  &~\,9.73  &~\,8.92  &  35.2  &  30.8  &  29.0  &  31.7  &  23.8  &  M5.0   \\
J1552-26  &  12.61  &  2.24  &  1.07  &  1.42  &  12.65  &  12.40  &  10.21  &~\,9.34  &  11.0  &  10.7  &~\,9.8  &  10.5  &~\,9.8  &  M6.0   \\
J1553-23  &  13.64  &  2.05  &  1.02  &  1.30  &  12.07  &  11.87  &~\,9.89  &~\,9.06  &  22.6  &  21.9  &  20.1  &  21.5  &  20.6  &  M5.0   \\
J1610-06  &  13.46  &  2.08  &  1.09  &  1.36  &  12.17  &  12.14  &  10.06  &~\,9.20  &  18.4  &  18.4  &  16.5  &  17.7  &  18.1  &  M5.5   \\
J1917-30  &  13.81  &  2.11  &  0.95  &  1.33  &  12.27  &  12.00  &~\,9.97  &~\,9.13  &  23.0  &  22.1  &  21.1  &  22.1  &  20.3  &  M5.5   \\
J2022-56  &  13.81  &  2.06  &  0.80  &  1.33  &  12.10  &  12.00  &~\,9.97  &~\,9.13  &  23.0  &  22.7  &  23.1  &  22.9  &  22.0  &  M5.5   \\
J2049-17  &  14.16  &  2.32  &  1.08  &  1.48  &  12.84  &  12.65  &  10.36  &~\,9.45  &  20.0  &  19.8  &  18.3  &  19.4  &  18.3  &  M6.0   \\
J2107-33  &  14.36  &  2.20  &  1.02  &  1.34  &  12.54  &  12.05  &  10.00  &~\,9.16  &  29.0  &  27.0  &  24.9  &  27.0  &  23.1  &  M5.5   \\
J2134-43  &  12.78  &  2.02  &  1.06  &  1.32  &  11.96  &  11.96  &~\,9.95  &~\,9.11  &  14.6  &  14.5  &  13.1  &  14.1  &  14.6  &  M5.5   \\
J2202-11  &  15.11  &  2.66  &  0.98  &  1.53  &  13.46  &  12.85  &  10.47  &~\,9.54  &  28.3  &  24.9  &  24.3  &  25.8  &  21.3  &  M6.5   \\
J2206-20  &  15.09  &  2.67  &  1.22  &  1.81  &  13.48  &  13.87  &  11.02  &~\,9.94  &  17.5  &  19.1  &  17.8  &  18.2  &  21.0  &  M8.0   \\
J2213-63  &  13.05  &  2.16  &  1.04  &  1.29  &  12.42  &  11.82  &~\,9.86  &~\,9.03  &  17.6  &  16.1  &  14.6  &  16.1  &  13.3  &  M5.0   \\
J2226-75  &  15.20  &  2.84  &  1.20  &  1.91  &  13.76  &  14.19  &  11.20  &  10.08  &  15.9  &  17.1  &  16.5  &  16.5  &  19.4  &  M8.5   \\
J2331-27  &  14.25  &  2.59  &  1.04  &  1.93  &  13.35  &  14.25  &  11.23  &  10.11  &  10.0  &  12.2  &  12.7  &  11.6  &  15.1  &  M8.5   \\
J2333-21  &  13.89  &  2.06  &  0.90  &  1.31  &  12.10  &  11.91  &~\,9.92  &~\,9.08  &  24.9  &  24.1  &  23.4  &  24.1  &  22.8  &  M5.0   \\
J2337-12$^{a}$&13.67&  2.13  &  1.10  &  1.36  &  12.33  &  12.14  &  10.06  &~\,9.20  &  20.2  &  19.8  &  17.7  &  19.2  &  18.5  &  M5.5   \\
    \noalign{\smallskip}
    \hline 
   \end{tabular}
  $$
  \begin{list}{}{}
  \item
$^{a}$: also discussed in Cruz \& Reid \cite{cruz} 

Column 1: Abreviated DENIS name.

Columns 2, 3 \& 4: DENIS photometry;
Column 5: PC3 index.

Column 6: Absolute $I$-band magnitude derived in Paper~II from the $I-J$
colour.

Column 7, 8 \& 9:  Absolute magnitudes for the $I$, $J$, $K$ bands 
based on the PC3-absolute magnitudes relation.

Columns 10, 11, 12: Distance (pc) estimated from the DENIS photometry and
the $M_{\rm I}$, $M_{\rm J}$, $M_{\rm K}$ derived from the PC3 index.

Columns 13 \& 14: Adopted distance, and distance previously derived in paper~II
from the the $I-J$ colour.

Column 15: Spectral types derived from the PC3 index, rounded to 
the nearest half subtype.

The star lists observed at T152 and NTT are merged and ordered by 
increasing RA.
  \end{list}
\end{table*}

As discussed above, the (PC3 vs. spectral type) relation of Mart\'{\i}n et al.
(\cite{martin99}) is our main spectral classification tool
for the M dwarfs, and we use the TiO5 vs. spectral type relation of 
Cruz \& Reid (\cite{cruz}) as a cross-check.
The two relations produce consistent spectral types, except beyond the
$\sim$M6 validity limit of TiO5. In Table~\ref{table_resultat} 
we therefore only list the PC3 spectral types. The Table
also lists the absolute magnitudes in the 3 DENIS bands computed from the
PC3 index, as well as the estimated spectrophotometric distances for 
each of those bands and their average. The values for the 3 colours $I$, 
$J$, $K$ are very similar, indicating correlated uncertainties for the 3
estimators. As customary, the distances and uncertainties assume that all
stars are single, while some fraction will in fact be unresolved binaries
and will have their distances underestimated by up to $\sqrt{2}$.
Table 3 with its improved distances contains 10 new stars closer than 15 pc, 
of which 2 are closer than 12 pc and one closer than 10 pc.

%
%
\section{Discussion}
\subsection{Distant reddened main sequence stars}
Three of the four reddened main sequence stars 
(DENIS-P J1552237$-$033520, DENIS-P J1553186$-$025919, and 
DENIS-P J1553251$-$044741) belong to our main sample, while 
DENIS-P J1625503$-$240008 = LP~862-26 is in the extended
sample.
%
All four stars passed our reduced proper motion selection (Paper II), and two 
have moderately high proper motions, 0.13 and 0.16 arc-sec.yr$^{-1}$. We 
expected both to be nearby M dwarfs, and their identification was a 
surprise to us, since we had restricted our
selection to high galactic latitudes ($|b_{II}|>$20 degrees) and
eliminated objects superimposed on large molecular complexes such as
Upper Scorpius. That selection turns out to have been insufficiently 
stringent, and all 4 objects are in the background of known high latitude 
molecular clouds, which had escaped our radar: J1552$-$03, J1553$-$04 
and J1553$-$02 are in the vicinity of the well known Lynds 134 and 
Lynds~183 (also known as Lynds~134N) dark nebulae,
and J1625$-$24 is in the background of the Rho Ophiucus molecular
complex. Our vetting procedure clearly needs to include smaller molecular
clouds than we had initially realised. Aside from this easily corrected 
shortcoming, our reduced proper motion selection proves very discriminating:
with an updated mask for molecular clouds, all observed targets match their
expected status.

\begin{figure}
\psfig{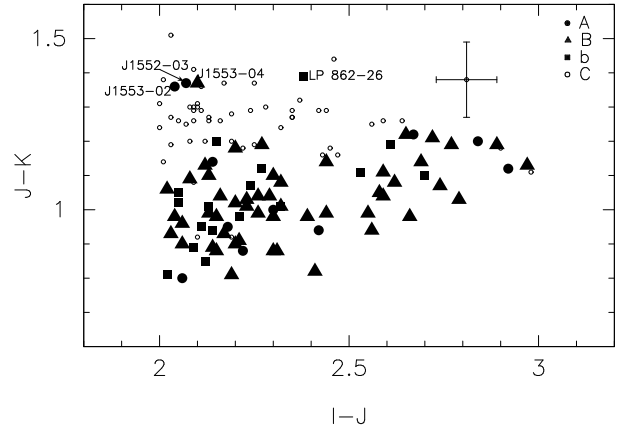}
\caption{\textit{DENIS} ($I-J, J-K$) colour-colour diagram of the 132 objects 
in the full sample. 
A, B and b groups = suspected M dwarfs, as defined in 
Sec.\ref{Data_sample};
C= suspected giants. 
The 4 objects at the top of the diagram are the strongly reddened
early type stars located behind molecular clouds; there are no other 
dwarf candidates in that part of the colour-colour diagram. The cross 
indicates representative error bars.}
\label{colcol.ps}
\end{figure}

Since we obtained spectra for only a fraction of the nearby star candidates,
one legitimate concern is whether the full sample contains additional 
reddened early-type contaminants. Fortunately, the ($I-J$, $J-K$) 
colour-colour diagram of the full sample (Figure~\ref{colcol.ps}) 
indicates that this is very unlikely. 
At a given $I-J$ colour, the giant candidates have a significantly larger 
$J-K$ colour than the dwarf candidates. At the precision of the DENIS 
photometry the two distributions overlap to some small extent around 
$I-J$=1.2, making that diagram an imperfect dwarf/giant diagnostic
tool (see Paper I, figure 2). The 
4 reddened stars, on the other hand, have even higher $J-K$ colours than 
any of the giants. They stand well clear of any 
of the nearby dwarf candidates, and we can thus safely conclude that 
the M dwarf lists of Papers~I and II contain no other reddened object.

\subsection{Low proper motion nearby M dwarfs}
After excluding the reddened early type stars, our main sample 
of 62 dwarf candidates still
contains 59 nearby dwarf candidates. 27 of them were observed, 
including all those with a proper motion below $0.1$~arc-sec.yr$^{-1}$.
As discussed above, it is unlikely that any of the unobserved stars
is a contaminant. Of those 59 dwarfs, 9 (15\%) have a proper motion 
below~ $ 0.1$~arc-sec.yr$^{-1}$ and a further 7 (11\%) between this value 
and the $0.18$~arc-sec.yr$^{-1}$ threshold of the 
NLTT, for a total of 26\%. This suggests that proper motion selection 
with traditional thresholds may miss up to a quarter of the 
d$<$30~pc population. Photometric binaries and Malmquist bias will
however both move some of those stars beyond 30~pc, making this a moderate
overestimate.

\begin{figure}
\psfig{file=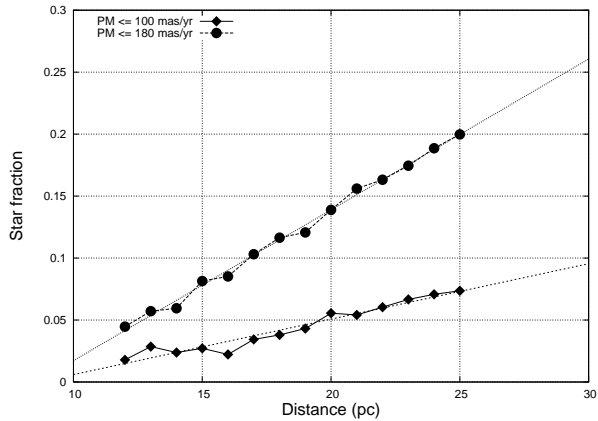,width=8.0cm,angle=-90}
\caption{Fractions of HIP stars brighter than $M_V = 9$ and with low PM 
(below 100 and 180 mas.yr$^{-1}$).  The HIP catalogue should be complete 
for them. The linear fits can be extrapolated up to 30 pc.}
\label{slowfrac} 
\end{figure}

Our main sample of A and B stars should be complete up to about 30 pc 
over the 5700 square degrees, as explained in Papers I and II.
In order to better evaluate the fraction of "slow stars" to be expected
versus distance, counts were performed in the HIPPARCOS Catalogue (ESA 1997). 
Of course it would have been possible to use some kinematical model with 
dispersions varying with the direction according to the velocity ellipsoid; 
but the counts are together very simple and quite accurate, and are 
therefore much more reliable and preferable.
{\it All} bright enough ($V \leq 11.5$), known or suspected, nearby 
stars were included in the Hipparcos Catalogue . For 25 pc this 
corresponds to $M_V = 9.5$ . On the other hand, the CNS3 (\cite{cns3}) is 
considered as complete to $M_V = 9$; therefore all stars closer than 25 pc and 
brighter than $M_V = 9$ should be in HIP and can be counted with their HIP 
distance and proper motion. 
Figure \ref{slowfrac} shows the fraction of HIP stars with PM smaller than 100
and 180 mas.yr$^{-1}$ and brighter than $M_V = 9$ vs distance. Although 
the correct analytical shape of these curves is not known with accuracy, 
clearly a linear fit can be made and safely extrapolated up to 30 pc.
The fractions of stars to be expected up to 30 pc with PM below 100 and 180 
mas.yr$^{-1}$ are respectively 10 and 26\%, not including the Poisson noise.
Our observed fraction of stars with PM below 100 mas.yr$^{-1}$ is slightly larger (15\%).

\section{Summary}

%
%

\begin{table}	
\caption{Sample summary}
\label{comptes}

 \begin{tabular}{c| c c c  c | c|c}

\hline
\hline Category, as  & A & B & A+B & C & b & Total \\
in Sec. 2.1 &  &  &  &  &  &\\
\hline Number of     & 11 & 51 & 62 & 52 & 18 & 132 \\ 
stars in  &  &  &  &  &  &\\
 Papers I \& II      &  &  &  &  &  &  \\
\hline Observed,     & 11 & 19 & 30 & 3 & 6  & 39 \\ 
this paper  &  &  &  &  &  &\\
\hline Classification  & 9Dw  & 18Dw  & 27Dw  & 3G & 5Dw   & 32Dw \\ 
after &  +2R & +1R & +3R &  & +1R & +4R\\
observation  &  &  &  &  &  & +3G \\ 

\hline

\end{tabular}  
\begin{trivlist}
\item

Columns A, B, (A+B) and C represent the contents of the main sample, over the
5700 square degrees;

Column "b" represents the extended sample (only NLTT stars), and is
not taken into account for the statistics.

Last line: "Dw" stands for "Dwarfs"; "R" for "reddened star"; "G" for "giant".
\end{trivlist}
\end{table}

Table \ref{comptes} summarizes the distribution of stars according to 
initial categories and final classification. Within the 5700 square 
degrees statistical sample, 3 of the 62 dwarf candidates 
are distant main sequence stars reddened by well known molecular 
clouds, and are easily recognisable from both their $J-K$ colour
and their position on the sky. With hindsight they could easily 
have been eliminated before spectroscopic observations. The others are all 
confirmed as late-type members of the immediate solar neighborhood. 9 of 
them have low proper motions, below 0.1~arc-sec.yr$^{-1}$. This 100\% success 
rate demonstrates that photometry and Reduced Proper Motion alone can be 
used to reliably identify members of the immediate solar neighbourhood. 
Spectroscopy obviously provides additional characterisation, but is not 
strictly needed for confirmation.

One quarter of the of 59 members of our statistically well defined sample 
of stars with a photometric distance below 30pc over 5700 square degrees 
has a proper motion below the NLTT threshold.

We calibrated the PC3 spectral index to absolute magnitudes, and used it 
to derive spectrophotometric distance to the 32~late-type M dwarfs. Three
objects have nominal distances within 12~pc: two new M8.5 dwarfs 
at 11~pc (DENIS-P~J0306$-$3647 and DENIS-P~J2331$-$2749), as well
as one previously known (but without distance) M7.5 dwarf at 10~pc 
(DENIS-P J1552446$-$262313, LP~860-41). 
\begin{acknowledgements}
The authors acknowledge help during the observations by the 2p2team 
at the European Southern Observatory, John Pritchard, Ivo Saviane and 
Rolando Vega. We also thank the DENIS consort for both data and good 
advice.The referee's comments were very useful for many improvements 
and clarifications.

P-B.N. acknowlegdes financial support from Vietnamese research grant 
B2004-09-13, and a nice Ph.D stay at CAI, Observatoire de Paris. 
E.M. acknowledges financial support from NSF grant AST 02-05862. 

This research has made use of the SIMBAD and VIZIER databases, 
operated at CDS, Strasbourg, France.
\end{acknowledgements}


\begin{thebibliography}{}

\bibitem[2002]{bakos}
  Bakos, G.A., Sahu, K.C., \& N\'emeth, P. 2002, ApJS, 141, 187

  
  
\bibitem[1990]{bessell} 
  Bessell, M.S. 1990, A\&AS, 83, 357


  
\bibitem[2002]{cruz}
  Cruz, K.L., \& Reid, I.N. 2002, AJ, 123, 2828

\bibitem[2002]{dahn02} 
  Dahn C.C., Harris H.C., et al. 2002, AJ, 124, 1170
  


\bibitem[1997]{delfosseb} 
  Delfosse, X. 1997, Ph.D Thesis, Grenoble University 
   
  
\bibitem[1997]{epchtein97} Epchtein, N. 1997, in 
the 2nd DENIS Euroconference,
The impact of large scale near-infrared surveys, 
ed. F. Garzon et al. (Kluwer Dordrecht), 15

\bibitem[1997]{esa}
  ESA 1997, The Hipparcos and Tycho Catalogues, ESA SP-1200
  
  
\bibitem[2002]{geballe}
  Geballe, T.R., et al. 2002, ApJ, 564, 466
  

\bibitem[Gliese \& Jahreiss, 1991]{cns3}
  Gliese,W., \& Jahreiss, H. 1991, Catalogue of Nearby Stars III, file 
available at CDS
  
  
  
  
\bibitem[1996]{hawley96}
  Hawley, S.L., Gizis, J.E., \& Reid, I.N. 1996, AJ, 112, 2799
  
\bibitem[2002]{hawley02}
  Hawley, S.L. et al. 2002, AJ, 123, 3409
  
    
  
  
        
  


 
\bibitem[1992]{leggett}
  Leggett, S.K. 1992, ApJS, 82, 531

\bibitem[1998]{leggett98}
  Leggett, S.K., Allard, F., \& Hauschildt, P.H. 1998, ApJ, 509, 836
  
\bibitem[2003]{lepine}
   L\'epine, S., Rich, R.M., \& Shara, M.M. 2003, AJ, 125, 1598
  
  

\bibitem[1979]{luytena} 
  Luyten, W.J. 1979, Catalogue of stars with proper motions exceeding 
   0\arcsec.5 annually (LHS) (Minneapolis, University of Minnesota)

\bibitem[1980]{luytenb} 
  Luyten, W.J. 1980, New Luyten catalog of stars with proper motions 
  larger than Two Tenths of an arcsecond (NLTT) (Minneapolis, University of Minnesota)
  
  


\bibitem[1999]{martin99} 
  Mart\'{\i}n, E. L., Delfosse, X., et al. 1999, AJ, 118, 2466

  
\bibitem[1992]{monet92}
  Monet, D.G., Dahn, C.C., et al. 1992, AJ, 103, 638
  
\bibitem[2003]{monet03}
  Monet, D.G., Levine, S.E., Canzian, B., et al. 2003, AJ, 125, 984
    
\bibitem[2001]{phan-bao01}
  Phan-Bao, N., Guibert, J., Crifo, F., Delfosse, X., Forveille, T., et al. 2001, A\&A, 380, 590
  
\bibitem[2002]{phan-bao02}
  Phan-Bao, N. 2002, Ph.D Thesis: {\it "Very low mass stars in the solar neighbourhood"}, Paris Observatory.
  
\bibitem[2003]{phan-bao03}
  Phan-Bao, N., Crifo, F., Delfosse, X., Forveille, T., Guibert, J., et al. 2003, A\&A, 401, 959

\bibitem[2005]{phan-bao05}
  Phan-Bao, N., Mart\'{\i}n, E.L., Reyl\'e, C., Forveille, T., \& Lim, J., 2005,
submitted to A\&A.

  
\bibitem[1995]{reid95} 
  Reid, I.N., Hawley, S.L., \& Gizis, J.E. 1995, AJ, 110, 1838


\bibitem[2003]{reid03}
  Reid, I.N. 2003, AJ, 126, 2449
  
  
\bibitem[2003]{salim}
  Salim, S., \& Gould, A. 2003, ApJ, 582, 1011
    

  
  
\bibitem[1993]{tinneya}
  Tinney, C.G., Mould, J.R., \& Reid, I.N. 1993, AJ, 105, 1045
  
\bibitem[1996]{tinney96} 
  Tinney, C.G. 1996, MNRAS, 281, 644 
  
  
\bibitem[1995]{vanaltena}
  van Altena, W.F., Lee, J.T., \& Hoffleit, E.D. 1995, The General Catalogue of Trigonometric
  Stellar Parallaxes, Fourth Edition (New Haven, CT: Yale University Observatory)
  
\bibitem[1996]{weis}
  Weis, E.W. 1996, AJ, 112, 2300 

\bibitem[1943]{vyss43} 
  Vyssotsky, A.N. 1943, Ap. J.,97, 381

\bibitem[1946]{vyss46} 
  Vyssotsky, A.N. 1946, Ap. J.,104, 234

\bibitem[1952]{vyss52} 
  Vyssotsky, A.N. 1952, Ap. J.,116, 117
 
\end{thebibliography}
\end{document}